\documentclass[conference]{IEEEtran}
\IEEEoverridecommandlockouts

\usepackage{cite}
\usepackage{amsmath,amssymb,amsfonts}
\usepackage{graphicx}
\usepackage{textcomp}
\usepackage{xcolor}
\usepackage{booktabs}
\usepackage{url} 
\usepackage{multirow}

\usepackage[hidelinks]{hyperref}

\DeclareRobustCommand*{\IEEEauthorrefmark}[1]{\raisebox{0pt}[0pt][0pt]{\textsuperscript{#1}}}

\def\BibTeX{{\rm B\kern-.05em{\sc i\kern-.025em b}\kern-.08em
    T\kern-.1667em\lower.7ex\hbox{E}\kern-.125emX}}

\begin{document}

\title{The SLT 2026 SmartGlasses Challenge: Benchmarking Egocentric Multi-Talker Speech Recognition and Understanding with Audio-Language Models}

\author{
    \IEEEauthorblockN{
        Dehui Gao\textsuperscript{1$\dagger$},
        Zhixian Zhao\textsuperscript{1$\dagger$},
        Zhennan Lin\IEEEauthorrefmark{1$\dagger$},
        Yujie Liao\IEEEauthorrefmark{1},
        Yuhang Dai\IEEEauthorrefmark{1},
        Yike Zhu\IEEEauthorrefmark{1},\\ 
        Longshuai Xiao\IEEEauthorrefmark{2},
        Hui Bu\IEEEauthorrefmark{3},
        Xin Xu\IEEEauthorrefmark{3}, 
        Xie Chen\IEEEauthorrefmark{4},
        Shuai Wang\IEEEauthorrefmark{5},
        Liumeng Xue\IEEEauthorrefmark{5},\\
        Zhonghua Fu\IEEEauthorrefmark{1},
        Jun Du\IEEEauthorrefmark{6},
        Eng-Siong Chng\IEEEauthorrefmark{7},
        Jun Zhou\IEEEauthorrefmark{8},
        Lei Xie\IEEEauthorrefmark{1*}
    }

    \vspace{0.15cm} 

    \IEEEauthorblockA{
        \IEEEauthorrefmark{1}ASLP@NPU, Northwestern Polytechnical University \\
        \IEEEauthorrefmark{2}Huawei \quad 
        \IEEEauthorrefmark{3}AIShell Inc \quad 
        \IEEEauthorrefmark{4}Shanghai Jiao Tong University \\\
        \IEEEauthorrefmark{5}Nanjing University \quad
        \IEEEauthorrefmark{6}University of Science and Technology of China\\
        \IEEEauthorrefmark{7}Nanyang Technological University \quad
        \IEEEauthorrefmark{8}Rokid
    }
}

\maketitle

\begingroup
\renewcommand\thefootnote{}
\footnote{$^\dagger$ These authors contributed equally to this work.}
\footnote{* Corresponding author.}
\addtocounter{footnote}{-2} 
\endgroup

\begin{abstract}
Recent advances in large language models (LLMs) and multimodal LLMs (MLLMs) have created new opportunities for wearable speech interfaces, with smart glasses providing an egocentric platform for continuous audio sensing and assistance. However, speech recognition and understanding in this setting remain challenging because of dynamic acoustic conditions, speaker overlap, and the spatial ambiguity introduced by wearer-centered recording geometry. To support systematic evaluation in this setting, we introduce the IEEE SLT 2026 SmartGlasses Challenge for egocentric multi-speaker speech processing. The challenge consists of two tracks, Dyadic Dialogue Understanding and Multi-party Meeting Understanding, and jointly evaluates Time-Stamped Speaker-Attributed Automatic Speech Recognition (TSA-ASR) and Spoken Language Understanding (SLU). It is built on a 106-hour four-channel egocentric speech dataset containing 714 sessions collected in real-world scenarios. This paper describes challenge tasks, dataset construction, submissions, and summarizes the main findings from the shared evaluation. The results show that heavy speaker overlap remains a major factor affecting TSA-ASR performance, while paralinguistic acoustic understanding continues to be difficult for current audio-language models in complex SLU settings. Further details can be found on the official challenge website\footnote{ \url{https://aslp-lab.github.io/SmartGlasses}}.
\end{abstract}


\section{Introduction}
\label{sec:intro}

Smart glasses are becoming an increasingly practical wearable interface for everyday AI assistance. Unlike smart speakers and smart phones, they can remain available throughout daily activities from a first-person perspective, making them well suited to spoken interaction and real-world audio-visual data capture. With recent advances in large language models (LLMs), multimodal large language models (MLLMs)~\cite{gpt4,llama3,qwen3omni,gemini31pro}, and audio-language models (ALMs)~\cite{salmonn,qwen_audio,whisper,qwen2_audio,audiopalm,speechgpt,pengi,ltu,audio_flamingo,kimiauudio}, smart glasses are becoming a more viable platform for speech-driven assistance and conversational support in realistic environments~\cite{project_aria}.

At the same time, speech processing on smart glasses remains difficult under real-world conditions. Because the device is wearable and mobile, it is used across diverse acoustic environments involving background noise, competing speakers, and variable interaction patterns~\cite{gin}. These challenges become more pronounced in egocentric multi-speaker speech recording, transcribing and understanding, where the wearer may participate either in dyadic dialogue or in longer multi-party meetings. Practical systems therefore need to handle speaker overlap, speaker attribution, temporal alignment, and long-context interpretation from continuously captured audio~\cite{peng2024speech_llm_survey,huo2026tagspeech}.

Existing corpora and benchmarks remain limited for this setting. Mainstream multi-talker and meeting datasets, such as AMI~\cite{ami}, ICSI~\cite{icsi}, AISHELL-4~\cite{aishell4}, AliMeeting~\cite{alimeeting}, and \mbox{NOTSOFAR-1}~\cite{notsofar1}, are primarily designed for fixed-position recording setups and do not reflect the variability of wearable egocentric audio.
Recent smart-glasses benchmarks, however, target different research goals and leave this joint evaluation setting unaddressed: WearVox~\cite{wearvox}, for instance, focuses on human-to-AI voice interaction, addressing tasks such as Side-Talk Rejection and Tool Calling, rather than dense multi-talker transcription, whereas CHiME-8 MMCSG~\cite{chime8_mmcsg} instead advances speaker-attributed speech recognition but does not extend to spoken language understanding. As a result, no existing benchmark provides a unified setup for jointly evaluating transcription and complex spoken language understanding in egocentric multi-speaker interaction~\cite{arora2024slue_perb}. Public benchmark coverage of Mandarin egocentric multi-speaker speech also remains limited.

In this paper, we present the IEEE SLT 2026 SmartGlasses Challenge, a Mandarin Chinese challenge on egocentric multi-speaker speech processing for wearable devices. The challenge is organized around two tracks, Dyadic Dialogue Understanding and Multi-party Meeting Understanding, and jointly evaluates Time-Stamped Speaker-Attributed Automatic Speech Recognition (TSA-ASR) and Spoken Language Understanding (SLU) under realistic acoustic conditions.

The main contributions of this paper are as follows:
\begin{itemize}
    \item We introduce the IEEE SLT 2026 SmartGlasses Challenge, which jointly evaluates TSA-ASR and SLU in Mandarin egocentric speech settings covering both dyadic dialogues and multi-party meetings.
    \item We present a four-channel egocentric speech corpus comprising 106 hours of audio and 714 sessions collected in realistic recording conditions.
    \item We summarize the participating systems and analyze the official evaluation results, highlighting the remaining challenges posed by heavy speaker overlap, long-context multi-speaker interaction, and SLU questions that cannot be resolved reliably from transcript information alone.
\end{itemize}

\section{Task Description}
\label{sec:task}

The challenge contains two tracks with different interaction settings but the same two sub-tasks, TSA-ASR and SLU. This design allows systems to be compared across scenario complexity while keeping the evaluation objectives consistent.

\subsection{Track 1: Dyadic Dialogue Understanding}
\label{ssec:track1}

Track 1 focuses on short face-to-face dyadic dialogues recorded in everyday environments. The dialogues are drawn from eight common real-world scenarios, including homes, restaurants, shopping malls, streets, and vehicle cabins, with an average duration of about 5.2 minutes per session. The main challenges in this track come from spontaneous overlap, background noise, frequent turn-taking, and context-dependent ellipsis in natural conversation.

\begin{itemize}
    \item \textbf{TSA-ASR Task:} Systems are required to generate speaker-attributed transcriptions with timestamps for dyadic dialogues under unseen background noise and speaker overlap. The main difficulty lies in assigning overlapped acoustic segments to the correct speakers. Performance is evaluated using tcpCER (Section~\ref{ssec:metrics}).
    
    \item \textbf{SLU Task:} Systems answer four-option multiple-choice questions (MCQs) about the recorded dialogue. The questions are balanced across acoustic reasoning, semantic reasoning, and acoustic-semantic joint reasoning. Performance is evaluated using Accuracy.
\end{itemize}

\subsection{Track 2: Multi-party Meeting Understanding}
\label{ssec:track2}

Track 2 focuses on longer meetings involving 3 to 8 participants. Compared with Track 1, this setting introduces more speakers, longer conversational context, denser overlap, and greater speaker-attribution difficulty, with an average duration of about 19 minutes per session. It is designed to test whether systems that perform well in short dyadic dialogues can generalize to more complex meeting-style interaction recorded from a wearable egocentric perspective.

\begin{itemize}
    \item \textbf{TSA-ASR Task:} Systems are required to perform speaker-attributed transcription with timestamps in multi-party meeting audio. Compared with Track 1, the task places greater demands on speaker tracking, time alignment, and robustness to dense overlap. Performance is evaluated using tcpCER.
    
    \item \textbf{SLU Task:} Systems answer MCQs about long meeting discussions. Compared with Track 1, the questions involve longer conversational context and place greater demands on long-context understanding, while still covering acoustic, semantic, and acoustic-semantic joint reasoning. Performance is evaluated using Accuracy.
\end{itemize}

\subsection{Evaluation Metrics}
\label{ssec:metrics}

TSA-ASR and SLU are evaluated with different primary metrics. Further details can be found here\footnote{\url{https://github.com/ASLP-lab/Smart-Glass-Challenge}}.

\textbf{TSA-ASR Metric:}
For TSA-ASR, we use Time-Constrained Polyphonic Character Error Rate (tcpCER) with a temporal collar of 5 seconds as the primary metric. The tcpCER is defined as:
\begin{equation}
\label{eq:tcpcer}
\mathrm{tcpCER} = \frac{\sum_{i=1}^{N}\left(S_i+D_i+I_i\right)}{\sum_{i=1}^{N}R_i} \times 100\%,
\end{equation}
where $N$ denotes the total number of reference turns; $S_i$, $D_i$, and $I_i$ denote the numbers of substituted, deleted, and inserted characters, respectively, in the $i$-th reference turn; and $R_i$ denotes the number of reference characters in that turn. Edit operations are computed after matching each hypothesis turn to its corresponding reference turn under the turn-constrained evaluation protocol~\cite{meeteval}.

\textbf{SLU Metric:}
For SLU, we use standard Accuracy as the primary metric, computed as the ratio of correctly answered multiple-choice questions (MCQs) to the total number of questions in the evaluation set.

\section{Dataset Construction}
\label{sec:dataset}

The dataset is constructed to support both speaker-attributed transcription and dialogue-level understanding from the same egocentric recordings. Its design therefore aims to preserve the acoustic variability of wearable recording conditions while maintaining annotation consistency and evaluation reliability. In this section, we describe the recording setup, collection protocol, annotation procedure, privacy and release policy, and overall dataset statistics.

\subsection{Recording Setup}
\label{ssec:setup}

\begin{figure}[t]
    \centering
    \includegraphics[width=0.6\linewidth]{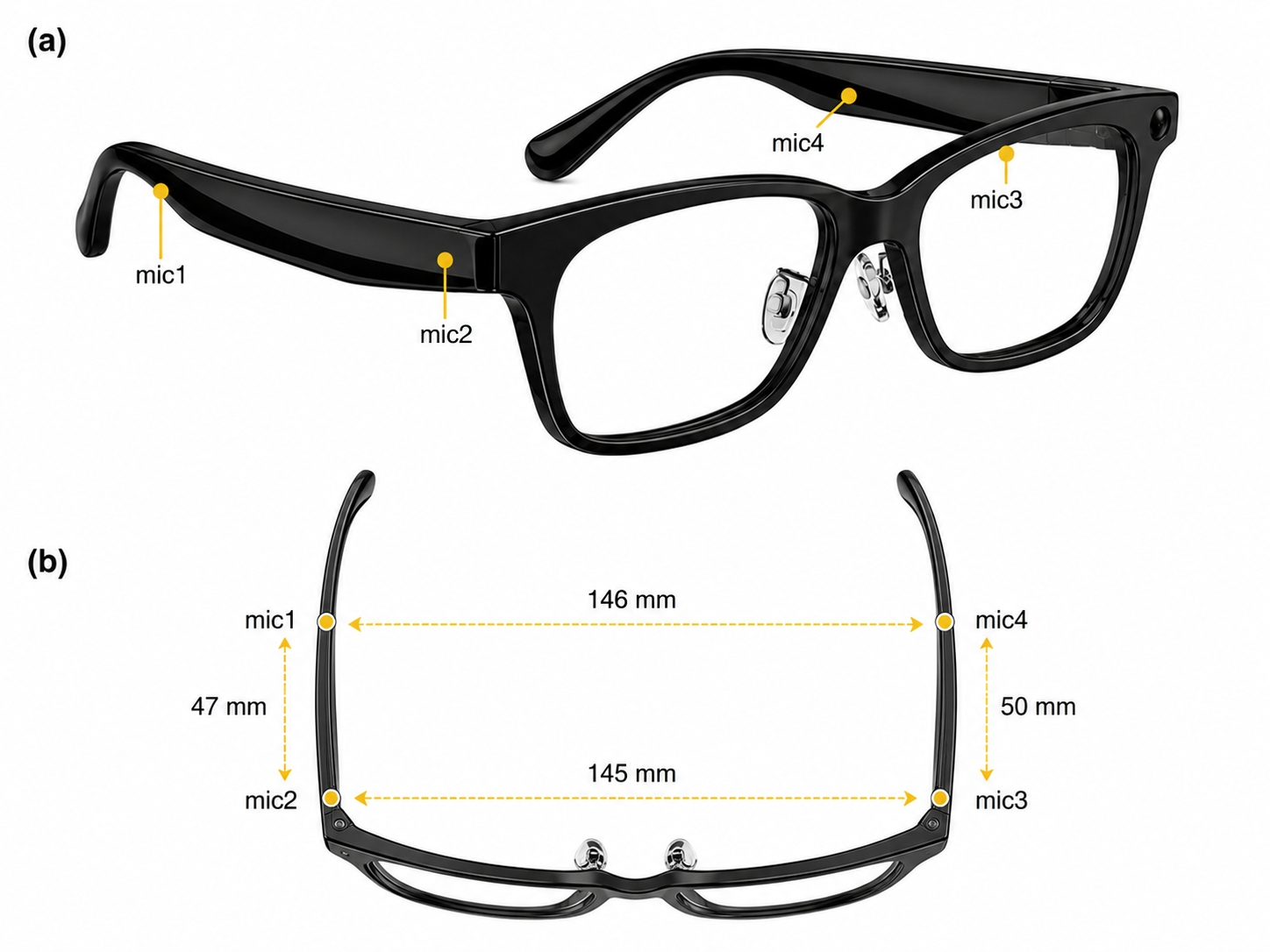}
    \caption{Hardware structure of the customized smart glasses and the spatial geometry of the four-channel microphone array. Microphones mic1--mic4 correspond directly to audio channels 1--4, respectively.}
    \label{fig:hardware}
\end{figure}

All recordings are collected using customized smart glasses equipped with a four-channel MEMS microphone array. The four microphones are mounted on the left and right temples, forming a fixed off-mouth wearable recording geometry, as illustrated in Fig.~\ref{fig:hardware}. Keeping the hardware configuration unchanged across sessions helps ensure that performance variation mainly reflects differences in scenario, speaker composition, and overlap condition rather than device inconsistency. All channels are synchronously recorded at 16~kHz with 16-bit precision.

\subsection{Data Collection}
\label{ssec:collection}

To balance conversational naturalness with evaluation needs, we adopt an ``Outline-Guided Spontaneous Conversation'' protocol. Fully scripted interactions would limit spontaneity, while unconstrained free conversation fails to guarantee the semantic complexity required for evaluation. We therefore use LLM-assisted conversation outlines~\cite{gemini31pro} that specify speaker roles, communicative goals, and key semantic events without prescribing exact utterances.

\begin{figure}[t]
    \centering
    \includegraphics[width=\linewidth]{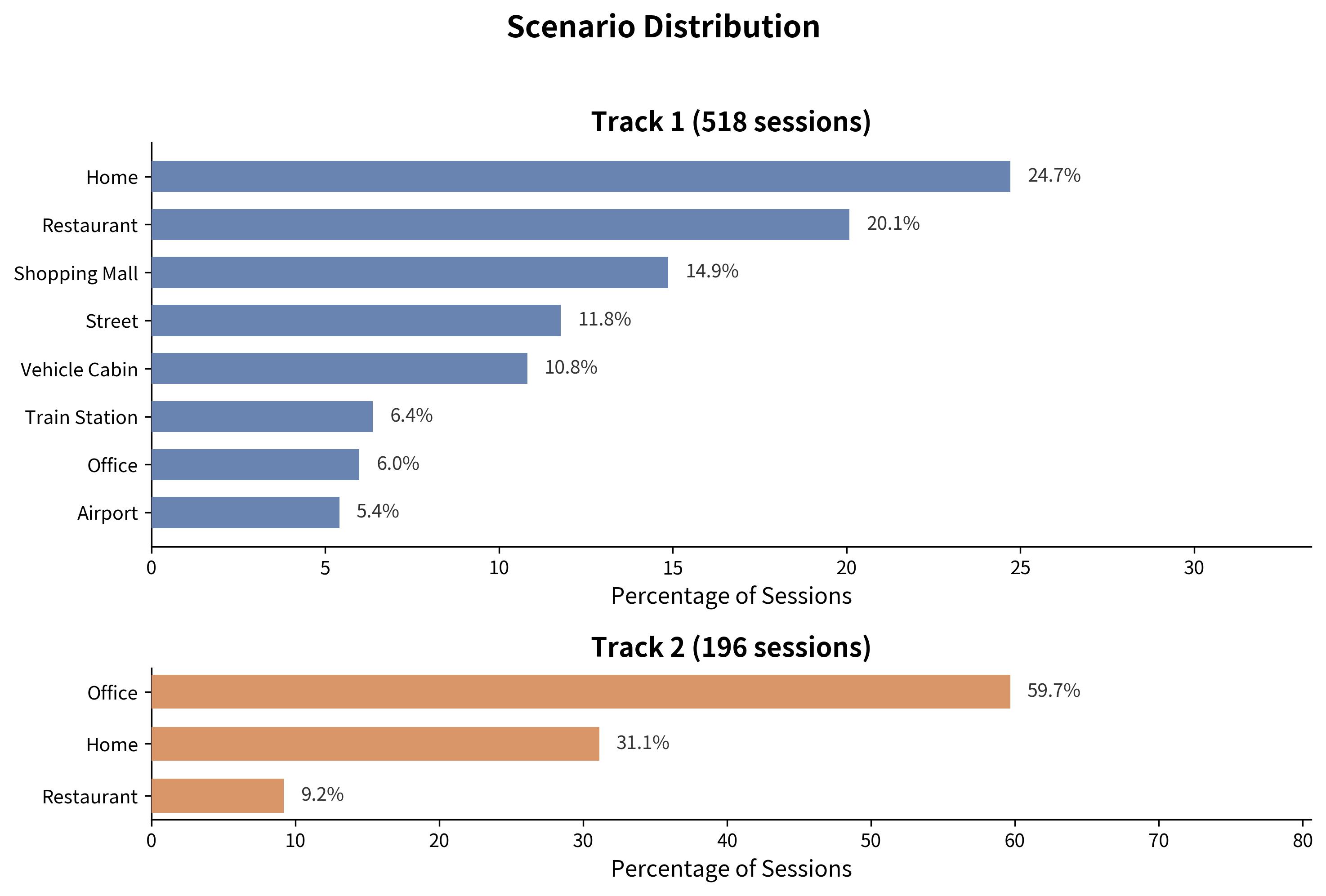}
    \caption{Proportional distribution of real-world acoustic scenarios recordings. The horizontal bar charts separately show the scenario composition for Track~1 (8 scenarios) and Track~2 (3 scenarios).}
    \label{fig:scenarios}
\end{figure}

Participants converse by following these outlines during recording. As a result, the sessions retain natural conversational phenomena such as interruptions, overlap, repetitions, and hesitations, while remaining suitable for downstream SLU question construction. The collected data cover eight daily scenarios for Track~1, including homes, restaurants, shopping malls, and streets, as well as three meeting-related scenarios for Track~2, as shown in Fig.~\ref{fig:scenarios}. All recordings are collected with informed consent, and only consented data are included in the released corpus.

\begin{table*}[t]
\centering
\caption{Taxonomy and Task Definitions of Spoken Language Understanding (SLU) Questions.}
\label{tab:slu_typology}
\begin{tabular}{lc p{5.5cm} p{6.5cm}}
\toprule
\textbf{Question Category} & \textbf{Ratio} & \textbf{Core Evidentiary Requirement} & \textbf{Standardized Task Subtypes} \\
\midrule
\textbf{Acoustic} & 1/3 & 
Requires direct reasoning over raw audio signals; unresolvable from text transcriptions alone. & 
Speaker Identification, Acoustic Event Detection, Speech Rate \& Prosody Analysis, Pitch/Volume Shift Tracking, Accent \& Dialect Profiling. \\
\addlinespace
\textbf{Semantic} & 1/3 & 
Primarily relies on textual/linguistic content derivable directly from speech transcriptions. & 
Dialogue State Tracking, Intent Classification, Key Information Retrieval, Causal \& Motivational Reasoning, Multi-Step Logical Deduction. \\
\addlinespace
\textbf{Acoustic-Semantic Joint} & 1/3 & 
Demands cross-modal alignment, integrating prosodic/acoustic cues with linguistic semantics. & 
Sarcasm \& Irony Detection, Paralinguistic Sentiment Analysis, Conversational Hesitation Analysis, Pragmatic vs. Literal Sense Disambiguation. \\
\bottomrule
\end{tabular}
\end{table*}

\subsection{Data Annotation}
\label{ssec:annotation}

The dataset is manually annotated in TextGrid format. For each utterance segment, annotators provide a pseudonymous speaker ID, timestamps, and the corresponding Mandarin Chinese transcription. All annotations are further checked through a two-pass cross-verification process.

For the SLU task, we construct four-option multiple-choice questions (MCQs) for both the development (Dev) and test (Test) sets. The question set is balanced across acoustic reasoning, semantic reasoning, and joint acoustic-semantic reasoning, following the 1:1:1 ratio shown in Table~\ref{tab:slu_typology}. Candidate question-answer pairs are first drafted with LLM assistance and then manually verified against the recorded dialogue. For the test set, each question undergoes two additional rounds of manual review to improve answer uniqueness and evidence consistency.

Before release, the corpus is de-identified by removing or masking direct personal identifiers in transcripts and metadata when present. Speaker labels are replaced with pseudonymous IDs, and only the information required for dataset description and evaluation is retained in the released version.

\subsection{Challenge Setup and Submission Constraints}
\label{ssec:rules}

To support fair comparison across submissions, the challenge adopts a shared set of constraints on model design, resource usage, and reproducibility. For TSA-ASR, submissions are expected to follow an end-to-end large-model paradigm for joint transcription and speaker attribution, while front-end auxiliary processing remains permitted. For SLU, reasoning is performed by an Audio-Language Model (ALM) that directly consumes audio, or audio-text, input rather than by a text-only cascade operating on ASR transcripts.

Separate systems may be used for TSA-ASR and SLU, although unified models are also allowed. External data, pre-trained models, and augmentation resources are limited to publicly accessible open-source materials and must be disclosed in the system description paper. The total learnable parameter count for a single-task system is limited to 35B, and top-performing submissions are verified through official reproduction in the final phase. Further details are available on the official challenge website\footnote{ \url{https://aslp-lab.github.io/SmartGlasses}}.




\subsection{Dataset Statistics}
\label{ssec:statistics}

The final dataset contains 106.98 hours of audio from 714 independent sessions. In total, 88 speakers (42 male and 46 female), aged 19 to 34, participate in the recordings. The session-level split is summarized in Table~\ref{tab:data_split}. Track~1 consists of 518 dyadic dialogues totaling 44.95 hours, while Track~2 contains 196 meetings totaling 62.03 hours. The number of participants in Track~2 ranges from 3 to 8. In addition, the evaluation sets include 3,509 SLU MCQs for fine-grained analysis of semantic and acoustic understanding.

\begin{figure}[t]
    \centering
    \includegraphics[width=0.9\linewidth]{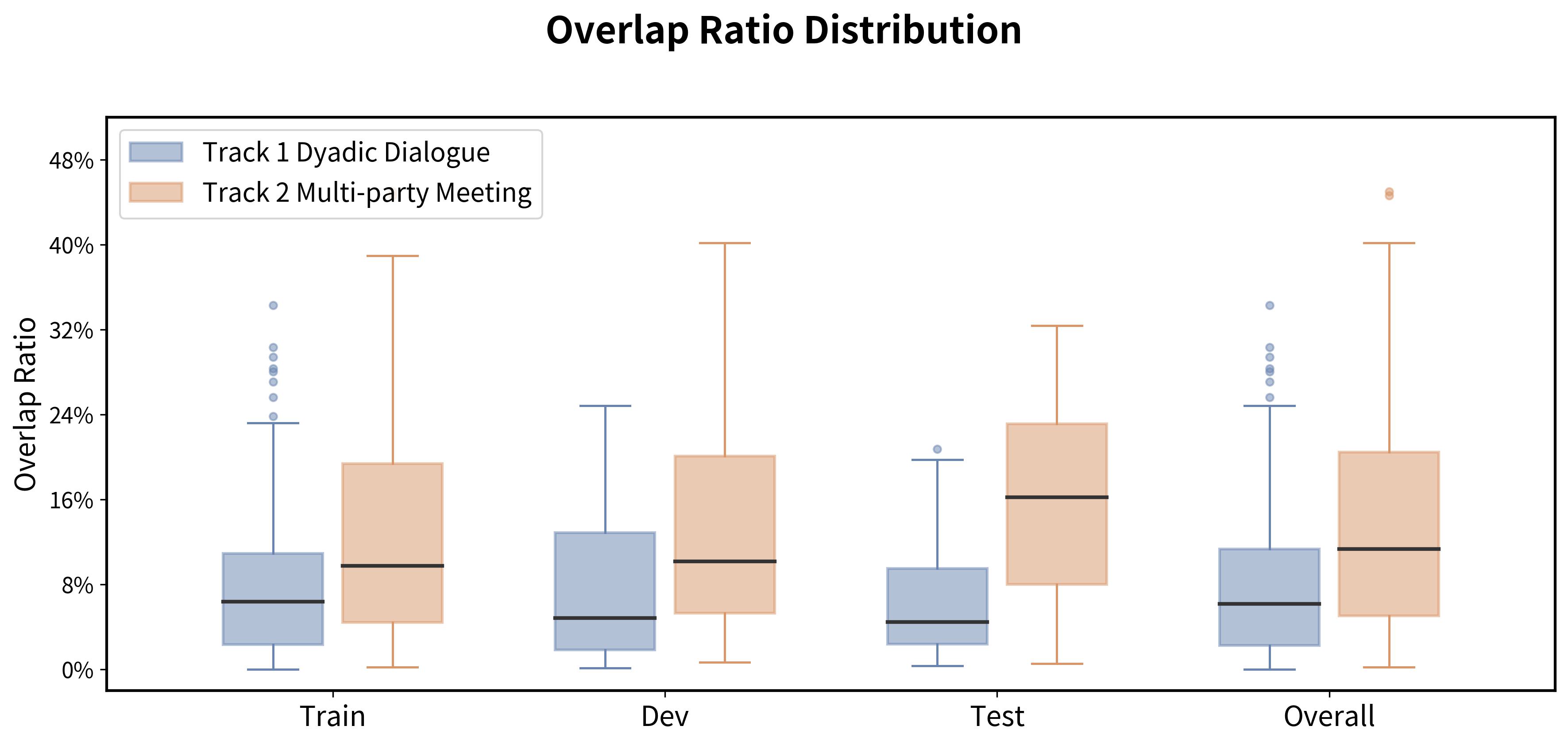}
    \caption{Distribution of speech overlap ratios across the Train, Dev, and Test splits, as well as the overall dataset. The box plots show the median, quartile ranges, and high-overlap outliers.}
    \label{fig:overlap}
\end{figure}

We also report speaker overlap statistics, as overlap is a major source of difficulty for speaker-attributed ASR in egocentric multi-speaker recordings. As shown in Fig.~\ref{fig:overlap}, the average overlap ratio is 7.5\% for Track~1 and 13.6\% for Track~2, while the maximum overlap ratio reaches approximately 35\% and 45\%, respectively, in high-overlap sessions.

\begin{figure}[t]
\vspace{2mm}
    \centering
    \includegraphics[width=0.8\linewidth]{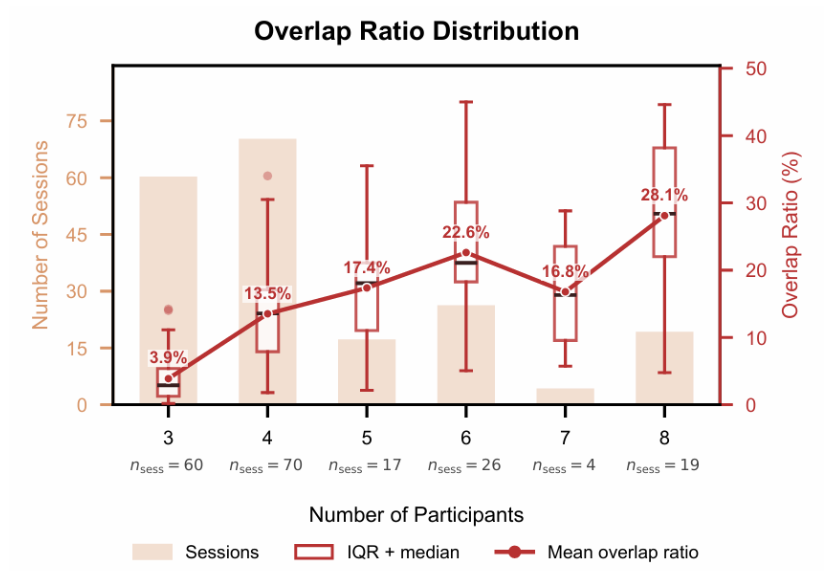}
    \caption{Track 2 overlap-ratio distributions by participant count across all splits. Bars show session counts, boxes show the interquartile range and median, and red markers and labels show group means. Here, $n_{\mathrm{sess}}$ denotes sessions per group and sums to 196.}
    \label{fig:track2_overlap_by_speakers}
\end{figure}

To further characterize the complexity of Track~2, we analyze overlap as a function of the number of meeting participants. As shown in Fig.~\ref{fig:track2_overlap_by_speakers}, Track~2 sessions span 3 to 8 participants, with most meetings involving 3 or 4 speakers. The mean overlap ratio generally increases with the number of participants, rising from 3.9\% in 3-speaker meetings to 28.1\% in 8-speaker meetings, although the trend is not strictly monotonic. Moreover, the distribution becomes more dispersed as the participant count increases, indicating greater variability in interaction patterns and a broader difficulty range in higher-cardinality meetings. These results suggest that the challenge of Track~2 comes not only from longer conversational context, but also from higher interaction density and speaker heterogeneity in multi-party scenarios.

\begin{table}[t]
\centering
\caption{Overall statistics and data split of the SLT 2026 SmartGlasses Challenge dataset.}
\label{tab:data_split}
\resizebox{\linewidth}{!}{
\begin{tabular}{lcccc}
\toprule
\textbf{Data Split} & \textbf{Sessions} & \textbf{Duration (h)} & \textbf{Avg. Len (s)} & \textbf{SLU QA Pairs} \\
\midrule
\multicolumn{5}{c}{\textit{Track 1: Dyadic Dialogue}} \\
\midrule
Train & 387 & 33.93 & 315.6 & - \\
Dev & 81 & 6.86 & 304.9 & 972 \\
Test & 50 & 4.16 & 299.8 & 588 \\
\textbf{Track 1 Total} & \textbf{518} & \textbf{44.95} & \textbf{312.4} & \textbf{1,560} \\
\midrule
\multicolumn{5}{c}{\textit{Track 2: Multi-party Meeting}} \\
\midrule
Train & 130 & 41.83 & 1158.3 & - \\
Dev & 36 & 11.51 & 1151.0 & 1068 \\
Test & 30 & 8.69 & 1042.2 & 881 \\
\textbf{Track 2 Total} & \textbf{196} & \textbf{62.03} & \textbf{1139.3} & \textbf{1,949} \\
\midrule
\textbf{Grand Total} & \textbf{714} & \textbf{106.98} & \textbf{-} & \textbf{3,509} \\
\bottomrule
\end{tabular}
}
\end{table}

\section{Results and Discussion}

In total, 15 teams made final submissions to the challenge, comprising 12 valid submissions for the TSA-ASR track and 9 valid submissions for the SLU track. The detailed performance and analysis for each task are presented as follows.

\subsection{Time-Stamped Speaker-Attributed ASR}
\label{ssec:asr_results}

Table~\ref{tab:tsa_asr} summarizes TSA-ASR performance on the two tracks. In Track~1, the top three systems all achieve tcpCER below 7\% on short dyadic dialogues, where the average overlap ratio is 7.5\%. These results suggest that, after adaptation to the challenge setting, several submitted systems can achieve relatively strong speaker-attributed transcription performance under moderate overlap in dyadic dialogues.

The comparison is markedly different in Track~2. This track contains long multi-party meetings with 3--8 speakers, an average duration of 19 minutes, and substantially denser overlap, with the maximum overlap ratio reaching 45\%. For systems evaluated on both tracks, performance degrades consistently from Track~1 to Track~2. The \textit{hfchen} team ranks first in Track~2 with a tcpCER of 27.95\%, while all other submitted systems remain above 48\%. This contrast indicates that long-form speaker-attributed transcription from an egocentric perspective remains substantially more difficult when dense overlap, longer conversational context, and frequent speaker switching are combined.


The submitted system reports reveal considerable diversity in backbone architecture, training strategy, and multi-channel audio processing. The following discussion summarizes the overall architectural landscape, highlights the design of the top-performing submission, and compares the different uses of four-channel audio. Since these observations are drawn from system descriptions rather than controlled ablation studies, they should not be interpreted as causal explanations for the observed performance.

\textit{Backbone architectures.} Consistent with the challenge rules, all valid submissions adopt end-to-end modeling, while differing in their backbone architectures and output formulations. The systems employ backbone models including SoulX-Transcriber~\cite{dai2026soulxtranscriber}, Qwen3-Omni~\cite{qwen3omni}, and VibeVoice-ASR~\cite{vibevoice_asr}. Other architectural choices include WavLM-based acoustic encoders~\cite{chen2022wavlm} and Mixture-of-Experts components for modeling complex and overlapping speech.

\textit{Design of the top-performing system.} The \textit{hfchen} system, ranked first on both tracks, adopts a MOSS Transcribe Diarize~(MTD)~\cite{mosstd} architecture that connects a Whisper-large-v3 acoustic encoder~\cite{whisper} to a Qwen3-8B decoder~\cite{qwen3} for unified prediction of speaker labels, timestamps, and transcriptions. It further performs continued audio-text pre-training on external multi-speaker data, with the aim of improving robustness to speaker changes, overlapping speech, and acoustic variability.

\textit{Diverse use of four-channel audio.} The submitted systems differ substantially in how they use the four-channel recordings. Some teams use a single channel, others average all four channels into a single waveform, and others investigate beamforming or separation for input construction while retaining end-to-end recognition. For example, \textit{kyousuke} evaluates direct four-channel averaging, GSS-based separation~\cite{gss}, and DPRNN\_mc-based multi-channel separation~\cite{luo2020dual}, but ultimately adopts channel averaging in its final submission.

Overall, the ASR results show that speaker overlap, long-context speaker tracking, and egocentric spatial variability remain central difficulties in this challenge setting. The leaderboard suggests that current systems are already relatively effective on moderate-overlap dyadic dialogues, but remain far less stable on long multi-speaker meetings recorded from a wearable egocentric perspective.

\subsection{Spoken Language Understanding}
\label{ssec:slu_results}

Table~\ref{tab:slu_results} presents the SLU accuracy results for both evaluation tracks. In Track~1, the top two systems achieve accuracy above 83\%, with the best-performing system reaching 88.8\%. In Track~2, the top three systems achieve accuracy above 84\%, and the best-performing system reaches 93.0\%. These results demonstrate the potential of current audio-language models for egocentric spoken language understanding. The performance variation across systems further suggests that substantial room remains for task-specific adaptation.

\textit{Backbone models and adaptation strategies.} Most SLU systems are built on Qwen-Omni variants~\cite{qwen2.5omni,qwen3omni}, although they differ in model scale and task adaptation. The top-performing \textit{hfchen} system instead adopts MOSS-Audio-8B-Thinking~\cite{openmoss2026mossaudio}, which combines a Qwen3-8B decoder~\cite{qwen3} with an audio encoder and supports reasoning-oriented generation. Beyond the backbone, \textit{jianfeng\_wu} relies more heavily on inference-time strategies such as adaptive audio cropping and voting-based prediction, while other systems emphasize additional question-answer construction, alignment-oriented tuning, and parameter-efficient adaptation such as QLoRA~\cite{dettmers2023qlora}.

\textit{Audio-text representation and fusion.} Another notable difference lies in how systems combine audio and text representations. The \textit{chenkang\_whu} system adopts a text-centered strategy, using textual representations as the main carrier for long-context reasoning while retaining audio input for paralinguistic cues. In contrast, systems such as \textit{jianfeng\_wu} and \textit{staragi} rely more directly on audio input, supported by techniques such as temporal cropping or acoustic representation repair. Their relative rankings vary across the two tracks, suggesting that the balance between audio and textual modalities remains sensitive to scenario difficulty and input length.

Overall, the SLU results indicate that modeling long-form audio context remains a notable challenge for egocentric speech understanding. While stronger backbone models and additional adaptation strategies are associated with better performance, achieving robust understanding in long multi-speaker scenarios continues to present substantial practical challenges within the current benchmark.

\begin{table}[t]
\centering
\caption{TSA-ASR performance (tcpCER \%) on dyadic dialogue (Track 1) and multiparty meeting (Track 2).}
\label{tab:tsa_asr}
\resizebox{\linewidth}{!}{
\begin{tabular}{lrcrc}
\toprule
\multirow{2}{*}{\textbf{Team}} & \multicolumn{2}{c}{\textbf{Track 1}} & \multicolumn{2}{c}{\textbf{Track 2}} \\
\cmidrule(lr){2-3} \cmidrule(lr){4-5}
& \textbf{Rank} & \textbf{tcpCER (\%)} & \textbf{Rank} & \textbf{tcpCER (\%)} \\
\midrule
hfchen        & 1  & 5.23  & 1 & 27.95 \\
kyousuke      & 2  & 6.22  & - & -     \\
xwzhang       & 3  & 6.57  & 2 & 48.92 \\
voxmindlabs   & 4  & 7.51  & 4 & 55.27 \\
chenkang\_whu & 5  & 8.72  & 6 & 59.05 \\
staragi       & 6  & 9.10  & 3 & 51.11 \\
jianfeng\_wu    & 7  & 9.27  & 7 & 95.80 \\
roysun2006    & 8  & 9.87  & - & -     \\
pcwang        & 9  & 10.57 & - & -     \\
shalcas       & 10  & 14.74 & - & -     \\
anyangfeng    & 11 & 27.44 & - & -     \\
Oracle        & -  & -     & 5 & 57.10 \\ \midrule
VibeVoice-ASR (Official Baseline)~\cite{vibevoice_asr} & - & 18.67 & - & 57.81 \\
\bottomrule
\end{tabular}
}
\end{table}

\begin{table}[t]
\centering
\caption{SLU performance (accuracy) on dyadic dialogue (Track 1) and multiparty meeting (Track 2).}
\label{tab:slu_results}
\resizebox{\linewidth}{!}{
\begin{tabular}{lrcrc}
\toprule
\multirow{2}{*}{\textbf{Team}} & \multicolumn{2}{c}{\textbf{Track 1}} & \multicolumn{2}{c}{\textbf{Track 2}} \\
\cmidrule(lr){2-3} \cmidrule(lr){4-5}
& \textbf{Rank} & \textbf{Accuracy} & \textbf{Rank} & \textbf{Accuracy} \\
\midrule
hfchen        & 1 & 0.888 & 1 & 0.930 \\
voxmindlabs   & 2 & 0.838 & 2 & 0.882 \\
jianfeng\_wu   & 3 & 0.808 & 4 & 0.812 \\
xwzhang       & 4 & 0.785 & 3 & 0.846 \\
pcwang        & 5 & 0.779 & - & -     \\
chenkang\_whu & 6 & 0.736 & 5 & 0.785 \\
kyousuke  & 7 & 0.728 & - & -     \\
shalcas       & 8 & 0.666 & - & -     \\
staragi       & 9 & 0.642 & 6 & 0.526 \\ \midrule
Qwen3-Omni-30B-A3B (Official Baseline)~\cite{qwen3omni} & - & 0.699 & - & 0.659 \\
\bottomrule
\end{tabular}
}
\end{table}

\subsection{Fine-grained Performance Analysis}
\label{ssec:fine_grained}

\begin{figure}[t]
    \vspace{2mm}
    \centering
    \includegraphics[width=\linewidth]{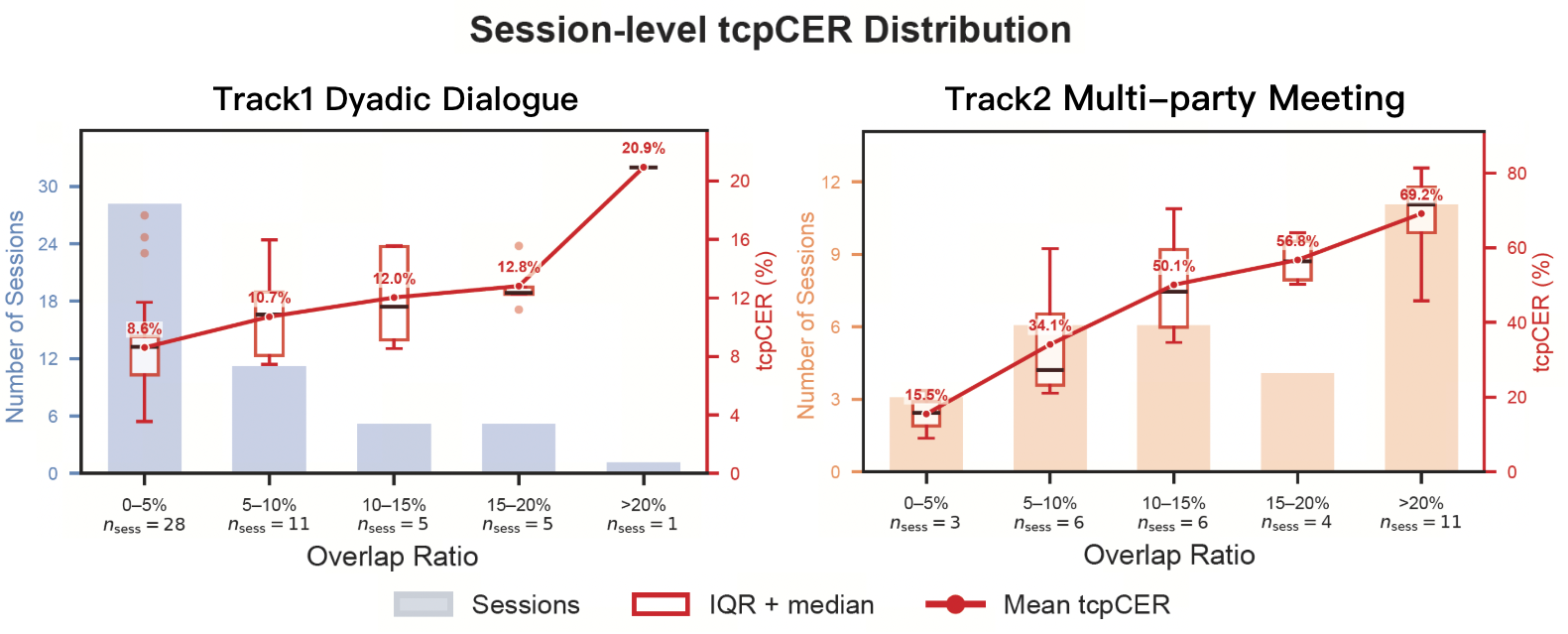}
    \caption{Test-set session-level tcpCER distributions by overlap interval after averaging the available submitted systems within each session. Bars show session counts, boxes show the interquartile range and median, and red markers and labels show interval means. Here, $n_{\mathrm{sess}}$ denotes test sessions per interval and sums to 50 and 30 for Tracks 1 and 2, respectively.}
    \label{fig:overlap_impact}
\end{figure}

\textit{Impact of Overlap Ratio on ASR Performance.}
To examine the effect of overlap more directly, we stratify session-level tcpCER by overlap ratio in Fig.~\ref{fig:overlap_impact}. In Track~1, the tcpCER distribution shifts gradually toward higher error rates as the overlap ratio increases, with the mean rising from 8.6\% in the 0--5\% interval to 14.2\% in the \textgreater{}15\% interval. The distribution remains relatively compact across groups. In Track~2, the distribution shifts upward more substantially and becomes more dispersed in higher-overlap intervals, reaching a mean of 69.2\% in the \textgreater{}20\% interval. These results are consistent with the leaderboard-level observation that overlap is a major factor affecting speaker-attributed ASR performance, especially in more complex multi-speaker settings.

\begin{figure}[t]
    \centering
    \includegraphics[width=0.6\linewidth]{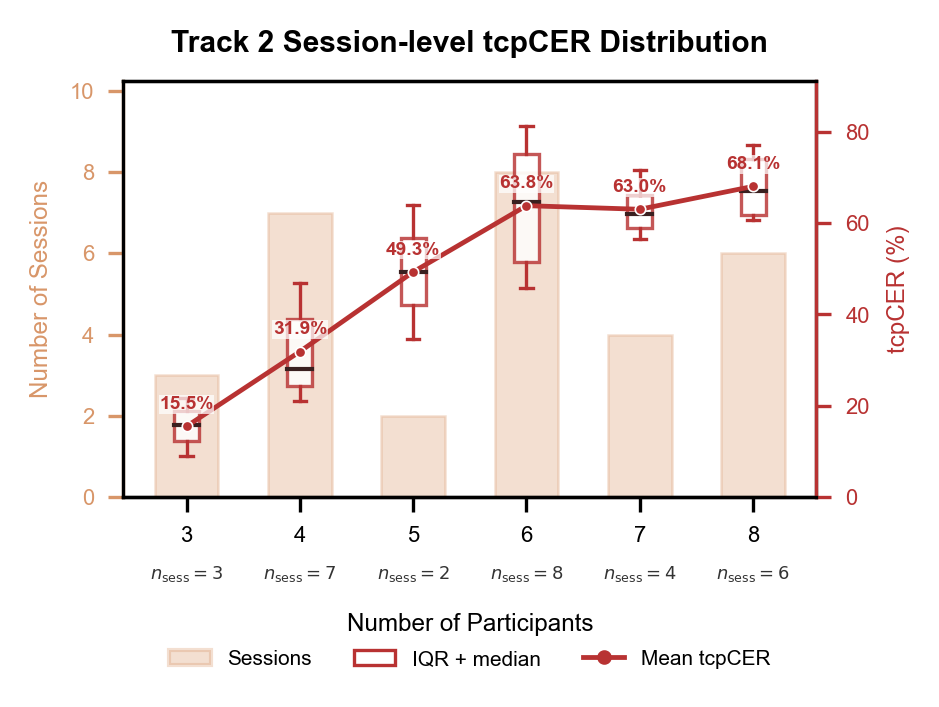}
    \caption{Track 2 test-set tcpCER distributions by participant count after averaging the available submitted systems within each session. Bars show session counts, boxes show the interquartile range and median, and red markers and labels show group means. Here, $n_{\mathrm{sess}}$ denotes test sessions per group and sums to 30.}
    \label{fig:participant_tcpcer}
\end{figure}

\textit{Impact of Number of Speakers on ASR Performance.}
To further characterize the difficulty of Track~2, we analyze session-level tcpCER on the test set as a function of the number of meeting participants in Fig.~\ref{fig:participant_tcpcer}. Although the number of sessions is uneven across participant-count groups, a clear upward trend can still be observed. The mean tcpCER rises from 15.5\% for 3-speaker meetings to 31.9\% for 4-speaker meetings and 49.3\% for 5-speaker meetings, and remains above 63\% for meetings with 6 to 8 participants. In addition to the increase in mean error, the tcpCER distribution becomes substantially wider in higher-cardinality meetings, with several high-error outliers. This pattern suggests that the difficulty of Track~2 is associated not only with overlap, but also with the increased complexity of multi-party interaction as the number of participants grows.

\begin{figure}[t]
    \vspace{2mm}
    \centering
    \includegraphics[width=\linewidth]{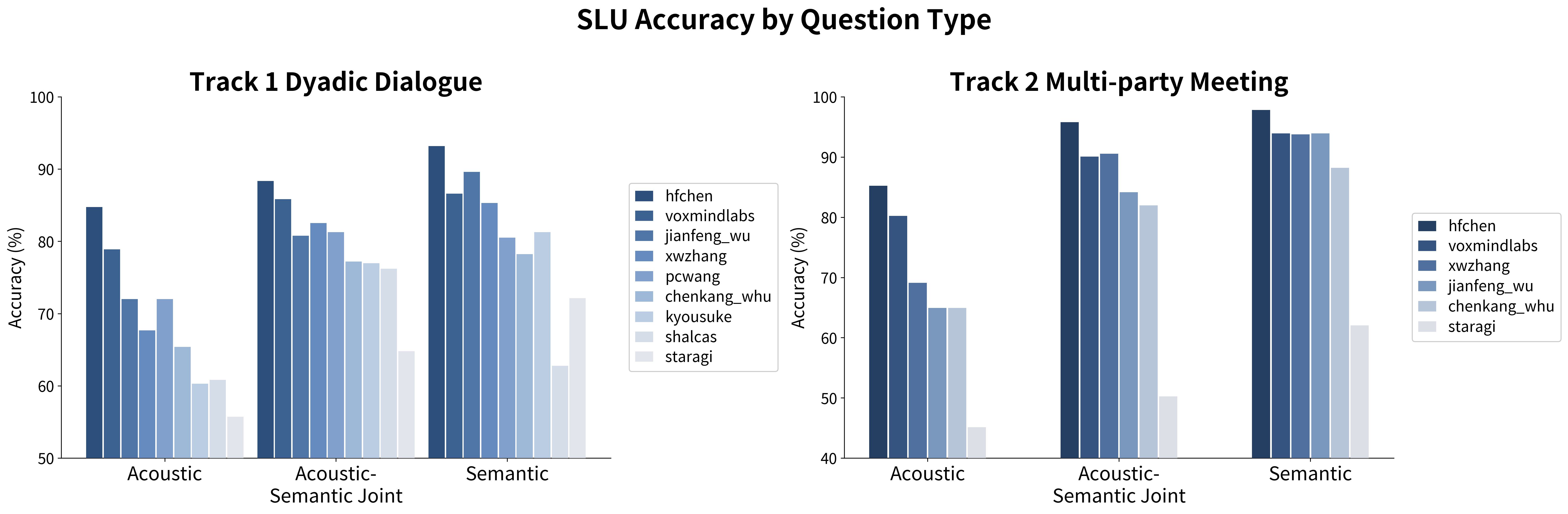}
    \caption{SLU accuracy breakdown across Acoustic, Acoustic-Semantic Joint, and Semantic question categories for top-performing teams.}
    \label{fig:slu_categories}
\end{figure}

\textit{Decoupling SLU Performance by Question Type.}
For the SLU task, we divide the test questions into Acoustic, Acoustic-Semantic Joint, and Semantic categories. As illustrated in Fig.~\ref{fig:slu_categories}, the evaluated systems generally achieve the highest accuracy on Semantic questions, lower accuracy on Joint questions, and the lowest accuracy on Acoustic questions. This trend is particularly visible in Track~2, where some systems exceed 90\% accuracy on Semantic questions but fall below 65\% on Acoustic questions. The gap suggests that current systems handle content-based understanding more robustly than questions that depend heavily on fine-grained acoustic evidence.

\subsection{Implications}
\label{ssec:implications}

The challenge results suggest that egocentric speech understanding should not be assessed by transcription quality alone. TSA-ASR and SLU reflect complementary aspects of performance, particularly in wearable multi-speaker settings where overlap, speaker attribution, and long-context reasoning interact. The results further indicate that long multi-party meetings remain substantially more difficult than short dyadic dialogues for current end-to-end systems.

\section{Conclusion}
\label{sec:conclusion}

This paper describes the SLT 2026 SmartGlasses Challenge for egocentric multi-speaker speech processing in wearable scenarios. The challenge combines a 106-hour four-channel speech corpus with paired TSA-ASR and SLU tasks spanning dyadic dialogues and multi-party meetings. The evaluation results highlight two recurring difficulties: TSA-ASR degrades substantially under dense overlap and increasing speaker complexity, while SLU remains weaker on questions requiring acoustic evidence than on primarily semantic questions. Collectively, these findings demonstrate that high transcription fidelity does not inherently guarantee robust speech understanding, underscoring several open directions for future work, including robust spatial modeling, long-context audio reasoning, and stronger acoustic-semantic grounding.


\bibliographystyle{IEEEtran}
\bibliography{main}

@article{gpt4,
  author        = {OpenAI and Josh Achiam and Steven Adler and Sandhini Agarwal and others},
  title         = {{{GPT-4}} Technical Report},
  journal       = {CoRR},
  volume        = {abs/2303.08774},
  year          = {2023},
}

@article{llama3,
  author        = {Aaron Grattafiori and Abhimanyu Dubey and Abhinav Jauhri and Abhinav Pandey and Abhishek Kadian and others},
  title         = {The {{Llama~3}} Herd of Models},
  journal       = {CoRR},
  volume        = {abs/2407.21783},
  year          = {2024},
}

@techreport{gemini31pro,
  author        = {{Google DeepMind}},
  title         = {{{Gemini 3.1 Pro}} Model Card},
  institution   = {Google DeepMind},
  address       = {London, U.K.},
  type          = {Model Card},
  year          = {2026},
  month         = feb,
  note          = {Accessed: Jul. 18, 2026. [Online]. Available: \url{https://storage.googleapis.com/deepmind-media/Model-Cards/Gemini-3-1-Pro-Model-Card.pdf}},
}

@inproceedings{salmonn,
  author        = {Changli Tang and Wenyi Yu and Guangzhi Sun and Xianzhao Chen and Tian Tan and others},
  title         = {{{SALMONN}}: Towards Generic Hearing Abilities for Large Language Models},
  booktitle     = {{ICLR}},
  year          = {2024},
}

@article{qwen_audio,
  author        = {Yunfei Chu and Jin Xu and Xiaohuan Zhou and Qian Yang and Shiliang Zhang and others},
  title         = {{{Qwen-Audio}}: Advancing Universal Audio Understanding via Unified Large-Scale Audio-Language Models},
  journal       = {CoRR},
  volume        = {abs/2311.07919},
  year          = {2023},
}

@article{qwen2_audio,
  author        = {Yunfei Chu and Jin Xu and Qian Yang and Haojie Wei and
                   Xipin Wei and Zhifang Guo and Yichong Leng and Yuanjun Lv and
                   Jinzheng He and Junyang Lin and others},
  title         = {{{Qwen2-Audio}} Technical Report},
  journal       = {CoRR},
  volume        = {abs/2407.10759},
  year          = {2024},
}

@inproceedings{whisper,
  author        = {Alec Radford and Jong Wook Kim and Tao Xu and
                   Greg Brockman and Christine McLeavey and Ilya Sutskever},
  title         = {Robust Speech Recognition via Large-Scale Weak Supervision},
  booktitle     = {Proceedings of the 40th International Conference on Machine Learning},
  series        = {Proceedings of Machine Learning Research},
  volume        = {202},
  pages         = {28492--28518},
  year          = {2023},
}

@article{audiopalm,
  author        = {Paul K. Rubenstein and Chulayuth Asawaroengchai and
                   Duc Dung Nguyen and Ankur Bapna and Zal{\'{a}}n Borsos and
                   Felix de Chaumont Quitry and Peter Chen and Dalia El Badawy
                   and Wei Han and Eugene Kharitonov and others},
  title         = {{{AudioPaLM}}: {A} Large Language Model That Can Speak and Listen},
  journal       = {CoRR},
  volume        = {abs/2306.12925},
  year          = {2023},
}

@article{speechgpt,
  author        = {Dong Zhang and Shimin Li and Xin Zhang and Jun Zhan and
                   Pengyu Wang and Yaqian Zhou and Xipeng Qiu},
  title         = {{{SpeechGPT}}: Empowering Large Language Models with Intrinsic
                   Cross-Modal Conversational Abilities},
  journal       = {CoRR},
  volume        = {abs/2305.11000},
  year          = {2023},
}

@inproceedings{pengi,
  author        = {Soham Deshmukh and Benjamin Elizalde and Rita Singh and
                   Huaming Wang},
  title         = {{{Pengi}}: An Audio Language Model for Audio Tasks},
  booktitle     = {Advances in Neural Information Processing Systems},
  volume        = {36},
  year          = {2023},
}

@inproceedings{ltu,
  author        = {Yuan Gong and Hongyin Luo and Alexander H. Liu and
                   Leonid Karlinsky and James Glass},
  title         = {Listen, Think, and Understand},
  booktitle     = {{ICLR}},
  year          = {2024},
}

@article{audio_flamingo,
  author        = {Zhifeng Kong and Arushi Goel and Rohan Badlani and
                   Wei Ping and Rafael Valle and Bryan Catanzaro},
  title         = {{{Audio Flamingo}}: {A} Novel Audio Language Model with
                   Few-Shot Learning and Dialogue Abilities},
  journal       = {CoRR},
  volume        = {abs/2402.01831},
  year          = {2024},
}

@inproceedings{ami,
  author        = {Jean Carletta and Simone Ashby and Sebastien Bourban and Mike Flynn and Mael Guillemot and others},
  title         = {The {{AMI}} Meeting Corpus: {A} Pre-announcement},
  booktitle     = {Machine Learning for Multimodal Interaction},
  series        = {Lecture Notes in Computer Science},
  volume        = {3869},
  publisher     = {Springer},
  pages         = {28--39},
  year          = {2006},
  doi           = {10.1007/11677482_3},
}

@inproceedings{icsi,
  author        = {Adam Janin and Don Baron and Jane Edwards and Dan Ellis and David Gelbart and others},
  title         = {The {{ICSI}} Meeting Corpus},
  booktitle     = {Proc. {ICASSP}},
  volume        = {1},
  pages         = {364--367},
  year          = {2003},
  doi           = {10.1109/ICASSP.2003.1198793},
}

@inproceedings{aishell4,
  author        = {Yihui Fu and Luyao Cheng and Shubo Lv and Yukai Jv and Yuxiang Kong and others},
  title         = {{{AISHELL-4}}: An Open Source Dataset for Speech Enhancement, Separation, Recognition and Speaker Diarization in Conference Scenario},
  booktitle     = {Proc. {Interspeech}},
  pages         = {3665--3669},
  year          = {2021},
  doi           = {10.21437/Interspeech.2021-1397},
}

@inproceedings{alimeeting,
  author        = {Fan Yu and Shiliang Zhang and Yihui Fu and Lei Xie and Siqi Zheng and others},
  title         = {{{M2MeT}}: The {{ICASSP}} 2022 Multi-Channel Multi-Party Meeting Transcription Challenge},
  booktitle     = {Proc. {ICASSP}},
  pages         = {6167--6171},
  year          = {2022},
  doi           = {10.1109/ICASSP43922.2022.9746465},
}

@inproceedings{notsofar1,
  author        = {Alon Vinnikov and Amir Ivry and Aviv Hurvitz and Igor Abramovski and Sharon Koubi and others},
  title         = {{{NOTSOFAR-1}} Challenge: New Datasets, Baseline, and Tasks for Distant Meeting Transcription},
  booktitle     = {Proc. {Interspeech}},
  pages         = {5003--5007},
  year          = {2024},
  doi           = {10.21437/Interspeech.2024-1788},
}

@article{project_aria,
  author        = {Zhaoyang Lv and Nicholas Charron and Pierre Moulon and Alexander Gamino and Cheng Peng and others},
  title         = {{{Aria}} Everyday Activities Dataset},
  journal       = {CoRR},
  volume        = {abs/2402.13349},
  year          = {2024},
}

@article{gin,
  author        = {Emilie d'Olne and Alastair H. Moore and Patrick A. Naylor and Thomas Lunner and others},
  title         = {Group Conversations in Noisy Environments ({{GiN}}) -- Multimedia Recordings for Location-Aware Speech Enhancement},
  journal       = {{IEEE} Open J. Signal Process.},
  volume        = {5},
  pages         = {374--382},
  year          = {2024},
  doi           = {10.1109/OJSP.2023.3344379},
}

@inproceedings{chime8_mmcsg,
  author    = {Katerina Zmolikova and Simone Merello and Kaustubh Kalgaonkar and Ju Lin and Niko Moritz and others},
  title     = {The {{CHiME-8}} {{MMCSG}} Challenge: Multi-modal Conversations in Smart Glasses},
  booktitle = {Proc. 8th International Workshop on Speech Processing in Everyday Environments (CHiME 2024)},
  pages     = {7--12},
  year      = {2024},
  doi       = {10.21437/CHiME.2024-2}
}

@inproceedings{wearvox,
  author        = {Zhaojiang Lin and Yong Xu and Kai Sun and Jing Zheng and Yin Huang and others},
  title         = {{{WearVox}}: An Egocentric Multichannel Voice Assistant Benchmark for Wearables},
  booktitle     = {{ICLR}},
  year          = {2026},
}

@article{qwen3omni,
  author  = {Jin Xu and Zhifang Guo and Hangrui Hu and Yunfei Chu and Xiong Wang and Jinzheng He and others},
  title   = {{{Qwen3-Omni}} Technical Report},
  journal = {CoRR},
  volume  = {abs/2509.17765},
  year    = {2025},
}

@article{qwen3,
  author  = {An Yang and Anfeng Li and Baosong Yang and Beichen Zhang and Binyuan Hui and others},
  title   = {{Qwen3} Technical Report},
  journal = {CoRR},
  volume  = {abs/2505.09388},
  year    = {2025},
}

@article{qwen2.5omni,
  author  = {Jin Xu and Zhifang Guo and Hangrui Hu and Yunfei Chu and Xiong Wang and Jinzheng He and others},
  title   = {{{Qwen2.5-Omni}} Technical Report},
  journal = {CoRR},
  volume  = {abs/2503.20215},
  year    = {2025},
}

@inproceedings{dettmers2023qlora,
  author    = {Tim Dettmers and Artidoro Pagnoni and Ari Holtzman and Luke Zettlemoyer},
  title     = {{{QLoRA}}: Efficient Finetuning of Quantized {LLMs}},
  booktitle = {Advances in Neural Information Processing Systems},
  volume    = {36},
  pages     = {10088--10115},
  year      = {2023},
}

@inproceedings{meeteval,
  author        = {Thilo von Neumann and Christoph Boeddeker and
                   Marc Delcroix and Reinhold Haeb{-}Umbach},
  title         = {{MeetEval:} {A} Toolkit for Computation of Word Error Rates
                   for Meeting Transcription Systems},
  booktitle     = {Proc. {CHiME} Workshop},
  pages         = {27--32},
  year          = {2023},
  doi           = {10.21437/CHiME.2023-6},
}

@article{chen2022wavlm,
  author  = {Sanyuan Chen and Chengyi Wang and Zhengyang Chen and Yu Wu and Shujie Liu and Zhuo Chen and Jinyu Li and Naoyuki Kanda and Takuya Yoshioka and Xiong Xiao and others},
  title   = {{{WavLM}}: Large-Scale Self-Supervised Pre-Training for Full Stack Speech Processing},
  journal = {{{IEEE}} J. Sel. Topics Signal Process.},
  volume  = {16},
  number  = {6},
  pages   = {1505--1518},
  year    = {2022},
}

@inproceedings{luo2020dual,
  author    = {Yi Luo and Zhuo Chen and Takuya Yoshioka},
  title     = {Dual-Path {RNN}: Efficient Long Sequence Modeling for Time-Domain Single-Channel Speech Separation},
  booktitle = {Proc. {ICASSP}},
  pages     = {46--50},
  year      = {2020},
}

@article{dai2026soulxtranscriber,
  author  = {Yuhang Dai and Haopeng Lin and Xinsheng Wang and others},
  title   = {{{SoulX-Transcriber}}: A Robust End-to-End Framework for Multi-Speaker Speech Transcription},
  journal = {CoRR},
  volume  = {abs/2606.02400},
  year    = {2026},
}

@article{mosstd,
  author  = {Mo Yu and Zheng-Yu Lin and Chen Yang and Yiyang Zhang and Hanfu Chen and others},
  title   = {{{MOSS}} Transcribe Diarize: Accurate Transcription with Speaker Diarization},
  journal = {CoRR},
  volume  = {abs/2601.01554},
  year    = {2026},
}

@article{openmoss2026mossaudio,
  author  = {{{OpenMOSS Team}}},
  title   = {{{MOSS-Audio}} Technical Report},
  journal = {CoRR},
  volume  = {abs/2606.01802},
  year    = {2026},
}

@inproceedings{gss,
  author        = {Desh Raj and Daniel Povey and Sanjeev Khudanpur},
  title         = {{GPU-accelerated} Guided Source Separation for Meeting
                   Transcription},
  booktitle     = {Proc. {Interspeech}},
  pages         = {3507--3511},
  year          = {2023},
  doi           = {10.21437/Interspeech.2023-42},
}

@article{vibevoice_asr,
  author        = {Zhiliang Peng and Jianwei Yu and Yaoyao Chang and
                   Zilong Wang and Li Dong and others},
  title         = {{{VIBEVOICE-ASR}} Technical Report},
  journal       = {CoRR},
  volume        = {abs/2601.18184},
  year          = {2026},
}

@article{kimiauudio,
  author        = {KimiTeam and Ding Ding and Zeqian Ju and Yichong Leng and
                   Songxiang Liu and Tong Liu and Zeyu Shang and Kai Shen and
                   Wei Song and Xu Tan and Heyi Tang and Zhengtao Wang and
                   Chu Wei and Yifei Xin and Xinran Xu and Jianwei Yu and
                   Yutao Zhang and Xinyu Zhou and others},
  title         = {{{Kimi-Audio}} Technical Report},
  journal       = {CoRR},
  volume        = {abs/2504.18425},
  year          = {2025},
}

@article{peng2024speech_llm_survey,
  author        = {Jing Peng and Yucheng Wang and Bohan Li and Yiwei Guo and
                   Hankun Wang and Yangui Fang and Yu Xi and Haoyu Li and
                   Xu Li and Ke Zhang and Shuai Wang and Kai Yu},
  title         = {A Survey on Speech Large Language Models for Understanding},
  journal       = {{IEEE} J. Sel. Topics Signal Process.},
  volume        = {20},
  number        = {1},
  pages         = {71--90},
  year          = {2025},
  doi           = {10.1109/JSTSP.2025.3546470},
}

@inproceedings{huo2026tagspeech,
  author        = {Mingyue Huo and Yiwen Shao and Yuheng Zhang},
  title         = {{{TagSpeech}}: End-to-End Multi-Speaker {ASR} and Diarization
                   with Fine-Grained Temporal Grounding},
  booktitle     = {Proceedings of the 64th Annual Meeting of the Association for Computational Linguistics},
  pages         = {41847--41862},
  year          = {2026},
}

@inproceedings{arora2024slue_perb,
  author        = {Siddhant Arora and Ankita Pasad and Chung-Ming Chien and
                   Jionghao Han and Roshan Sharma and Jee-weon Jung and
                   Hira Dhamyal and William Chen and Suwon Shon and
                   Hung-yi Lee and Karen Livescu and Shinji Watanabe},
  title         = {On the Evaluation of Speech Foundation Models for Spoken
                   Language Understanding},
  booktitle     = {Proc. {ACL} Findings},
  year          = {2024},
}


\end{document}